Global network structure of dominance hierarchy of ant workers

Hiroyuki Shimoji[1,2,†,‡], Masato S. Abe[2,†], Kazuki Tsuji[1], Naoki Masuda[3,4*]

1. Department of Agro-Environmental Sciences, Faculty of Agriculture, University of the Ryukyus, Okinawa 903-0213, Japan

2. Department of General Systems Studies, Graduate School of Arts and Sciences, University of Tokyo, Tokyo 153-8902, Japan

3. Department of Engineering Mathematics, University of Bristol, Bristol, BS8 1UB, United Kingdom

4. CREST, JST, Kawaguchi, Saitama, 332-0012, Japan

† Contributed equally to this work

* Author for correspondence (N. Masuda: naoki.masuda@bristol.ac.uk)

‡ Current address: Laboratory of Ecological Genetics, Graduate School of Environmental Science, Hokkaido University, Sapporo, Hokkaido, 060-0810, Japan

**Abstract**

Dominance hierarchy among animals is widespread in various species and believed to serve to regulate resource allocation within an animal group. Unlike small groups, however, detection and quantification of linear hierarchy in large groups of animals are a difficult task. Here, we analyse aggression-based dominance hierarchies formed by worker ants in *Diacamma* sp. as large directed networks. We show that the observed dominance networks are perfect or approximate directed acyclic graphs, which are consistent with perfect linear hierarchy. The observed networks are also sparse and random but significantly different from networks generated through thinning of the perfect linear tournament (i.e., all individuals are linearly ranked and dominance relationship exists between every pair of individuals). These results pertain to global structure of the networks, which contrasts with the previous studies inspecting frequencies of different types of triads. In addition, the distribution of the out-degree (i.e., number of workers that the focal worker attacks), not in-degree (i.e., number of workers that attack the focal worker), of each observed network is right-skewed. Those having excessively large out-degrees are located near the top, but not the top, of the hierarchy. We also discuss evolutionary implications of the discovered properties of dominance networks.

# 1. Introduction

Dominance interaction such as aggressive physical interaction and ritualised displays between dominant (i.e., high-ranked in the hierarchy) and subordinate (i.e., low-ranked) individuals is widespread in animals. Dominance hierarchies are a regulatory mechanism of the social system and observed in a wide range of taxa from vertebrates to invertebrates [1–4]. Dominant individuals would have a high chance to access to resources and can enhance their fitness, whereas subordinates have a small chance of resource acquisition and as a result in some taxa undertake a role of helper that does not reproduce [5, 6].

Eusocial insects characterised by reproductive division of labour provide opportunities for studying complex social organisations [1, 7, 8]. In many eusocial Hymenoptera (e.g., ants, honey bees, and wasps), workers cannot mate but do produce males through arrhenotokous parthenogenesis. Kin selection theory suggests that colony members are in conflict over male production because a worker gains genetic benefit by rearing her own sons [9–11]. At the same time, worker reproduction is costly to a colony because reproductive workers allocate their workforces to personal reproduction rather than to household chores for maintaining the colony [12, 13] (see also [8]). In fact, most workers facing this conflict remain sterile. Dominance interaction is also found in various species of ants and regulates worker reproduction [14–18].

A prevalent theoretical approach to dominance hierarchy is to rank individuals in a group on the basis of observed dyadic interactions. By inferring the direction of missing interactions between pairs of individuals if necessary, one can construct a so-called tournament, which is an assignment of directed links to all pairs of individuals [19, 20]. If cyclic dominance relationship (e.g., rock-scissors-paper relationship among three individuals) is absent, the linear ordering uniquely exists such that a dominant individual always has a higher rank than the subordinate individual in any pair [21, 22]. The degree of linear hierarchy is related to relative abilities of individuals in controlling resources such as mates and food [4] and to group stability [23, 24].

However, linear hierarchy is commonly violated in large groups of animals in various species [20–22, 25, 26]. In particular, pair-flips, i.e., bidirectional links [24], and intransitive triads represented by the rock-scissors-paper relationship [22, 25, 26] are basic building blocks that make dominance networks not linear. The loss of linearity is intuitive given that an individual would not be able to recognise all peers and many individuals would have similar strengths in a large group.

In the present study, we observe aggressive behaviour among workers of an ant species *Diacamma* sp. on a large scale (i.e., 58–214 workers). We show that the observed dominance networks are directed acyclic graphs (DAGs) or approximately DAGs such that they are consistent with (almost) perfect linear hierarchy, despite the relatively large group sizes. Furthermore, using network analysis tools specialised for directed networks, particularly those recently developed for DAGs, we analyse rank-dependent aggression behaviour of individuals and randomness inherent in global structure of the observed networks.

## 2. Materials and methods

### 2.1 Ant species

*Diacamma* sp. is a ponerine species. In Japan, this is the only species of the genus *Diacamma*. Colonies are monogynous containing at most one functional queen in each colony together with 20–300 workers [27]. Precisely speaking, queens of this species are called gamergates, i.e., mated reproductive workers [28], because unlike many other ants the role differentiation between queens and workers in this species occurs not through the larval development but via the specialised social manipulation after the adult eclosion called the gemmae mutilation [29]. In the current study, we use the terms queen and worker for simplicity. The queen monopolises the production of female offspring. Workers, i.e., those whose gemmae are mutilated, cannot mate but retain the ability to produce male-destined haploid eggs. Aggressive behaviours are frequently observed among workers when the queen is absent or the colony is large [27, 30]. Such behaviour is considered to reflect competition over direct male production, because the dominant workers, usually the most dominant one, actually lay eggs even in the presence of the queen especially in large colonies [27, 31]. The queen never participates in dominance hierarchy.

There is another type of aggressive interaction among nestmate workers in this species, i.e., worker policing [32]. Worker policing is behaviourally distinct from aggressive behaviour underlying dominance hierarchy. In worker policing, multiple individuals simultaneously attack one victim to immobilise it. In contrast, aggression takes the form of one-on-one interaction, i.e., biting and jerking [30], which has led us to study the dominance hierarchy by network analysis tools; a network is by definition composed of a collection of pairwise interactions.

We collected six colonies of *Diacamma* sp. in peninsula Motobu, northern part of the main island of Okinawa, in March-May 2011 and in August 2012. Each colony contained a queen and 58–214 workers. We individually marked all the adult ants by enamel markers. Then, we housed each colony in "double container" artificial nests [27], which comprised a small plastic container (10 cm × 10 cm × 2.5 cm high) serving as a breeding chamber. The container was located in the centre of a larger container (15 cm × 21 cm × 13.5 cm high). The colonies had been maintained in the laboratory (25 ± 1 °C, 14 L: 10 D cycle) for two to three months before the observation began. The workers were fed ad libitum on honey worms (caterpillars of *Galleria mellonella*) and water placed outside the smaller container. It should be noted that we recorded aggressive behaviour among workers in the presence of the queen in each colony.

**2.2 Recording aggressive behaviour**

We recorded all aggressive behaviour events between all pairs of workers in each colony for 300 min per day for consecutive four days using a digital video camera (HDR-CX700V, Sony, Japan). The aggressive interaction was defined as bite and jerk [30]. We counted the number of aggressive interactions between each pair of workers and constructed a directed social network for each colony. The nodes are workers. A directed link represents a pair of workers that have interacted at least once and emanates from the attacking worker to the attacked worker.

**2.3 Triangle transitivity metric**

Due to high sparseness of the recorded networks (see Results), we should resort to a measurement different from well-known indices such as Landau's $h$ and Kendall's $K$ to quantify linearity (i.e., orderliness) of a dominance network. To this end, we calculate the triangle transitivity metric [22]. The triangle transitivity is defined as a normalised value of the number of transitive triads (A attacks B, A attacks C, and B attacks C) divided by the sum of the number of transitive triads and that of cycles (A attacks B, B attacks C, and C attacks A). See Electronic Supplementary Material for the mathematical definition.

**2.4 Generation of thinned linear tournaments**

We generate a thinned linear tournament possessing $N$ nodes and $|E|$ expected number of links, which is used as a null model for probing structure of observed networks, as follows (equivalent to the cascade model used in food web research [33]). Consider the linear tournament with $N$ nodes, in which all pairs of individuals interact (i.e., complete graph) and perfectly ranked in the sense that the higher-ranked individual dominates the lower-ranked individual in any pair [1, 34]. Then, we independently retain each of the $N(N-1)/2$ directed links with probability $p$. Otherwise, we remove the link. We call the generated network a thinned linear tournament. We set

$$p = \frac{2|E|}{N(N-1)} \quad (1)$$

such that the expected number of the links in the generated network is equal to $|E|$.

**2.5 CV**

The in-degree of a worker is defined as the number of directed links incoming to the worker, i.e., number of workers that attack the focal worker. The out-degree of a worker is defined as the number of directed links outgoing from the worker, i.e., number of workers that the focal worker attacks. If we convert the bidirectional links to unidirectional links by discarding one of the two directions whose link weight is smaller than the other (Electronic Supplementary Material, section S.4), the out-degree is identical to the *Netto* dominance index [35].

To quantify the heterogeneity in the in- and out-degree, we measure the coefficient of variation (CV), i.e., the ratio of the standard deviation of in- or out-degree to the mean. If all workers have the same degree, CV is equal to zero. The exponential degree distribution yields CV equal to unity. A CV value much larger than unity implies that the distribution is heavily tailed.

**2.6 CV for the thinned linear tournament**

In the thinned linear tournament, the probability that the worker with rank $k$ $(1 \le k \le N)$ has in-degree $d^{\text{in}}$ $(0 \le d^{\text{in}} \le k-1)$ is given by

$$P(d^{\text{in}} | k) = \frac{(k-1)!}{d^{\text{in}}!(k-1-d^{\text{in}})!} p^{d^{\text{in}}} (1-p)^{k-1-d^{\text{in}}}. \tag{2}$$

Therefore, the in-degree distribution not conditioned by the rank is given by

$$\begin{aligned} P(d^{\text{in}}) &= \frac{1}{N} \sum_{k=d^{\text{in}}+1}^{N} P(d^{\text{in}} | k) \\ &= \frac{1}{N} \sum_{k=d^{\text{in}}+1}^{N} \frac{(k-1)!}{d^{\text{in}}!(k-1-d^{\text{in}})!} p^{d^{\text{in}}} (1-p)^{k-1-d^{\text{in}}}. \end{aligned} \tag{3}$$

Similarly, the conditional and unconditional distributions of the out-degree $d^{\text{out}}$ $(0 \le d^{\text{out}} \le N-k)$ are given by

$$P(d^{\text{out}} | k) = \frac{(N-k)!}{d^{\text{out}}!(N-k-d^{\text{out}})!} p^{d^{\text{out}}} (1-p)^{N-k-d^{\text{out}}} \tag{4}$$

and

$$\begin{aligned} P(d^{\text{out}}) &= \frac{1}{N} \sum_{k=1}^{N-d^{\text{out}}} P(d^{\text{out}} | k) \\ &= \frac{1}{N} \sum_{k=1}^{N-d^{\text{out}}} \frac{(N-k)!}{d^{\text{out}}!(N-k-d^{\text{out}})!} p^{d^{\text{out}}} (1-p)^{N-k-d^{\text{out}}} \\ &= \frac{1}{N} \sum_{k'=d^{\text{out}}+1}^{N} \frac{(k'-1)!}{d^{\text{out}}!(k'-1-d^{\text{out}})!} p^{d^{\text{out}}} (1-p)^{k'-1-d^{\text{out}}}, \end{aligned} \tag{5}$$

respectively. Because the unconditional in- and out-degree distributions are the same, the expected in-degree and out-degree are given by

$$E(d^{\text{in}}) = E(d^{\text{out}}) = \sum_{d=0}^{N-1} dP(d) = \frac{(N-1)p}{2}. \tag{6}$$

The variance of the in- and out-degree is given by

$$V(d^{\text{in}}) = V(d^{\text{out}}) = \sum_{d=0}^{N-1} P(d) \left[ d - \frac{(N-1)p}{2} \right]^2 = \frac{(N-1)p[(N-5)p+6]}{12}. \tag{7}$$

By combining equations (1), (6), and (7), we obtain the CV for the thinned linear tournament as follows:

$$\text{CV} = \frac{\sqrt{V(d^{\text{in}})}}{E(d^{\text{in}})} = \frac{\sqrt{V(d^{\text{out}})}}{E(d^{\text{out}})} = \sqrt{\frac{|E|(N-5) + 3N(N-1)}{3|E|(N-1)}}. \tag{8}$$

**2.7 Bottom-up leaf-removal algorithm and ranking of workers**

We determine the ranks of nodes (i.e., workers) in a DAG using the so-called bottom-up leaf-removal algorithm [36, 37] as follows (see Figure S1 for a schematic). First, we remove the nodes without outgoing links. The set of the removed nodes is denoted by $W_1$. We also remove the incoming links incident to the removed nodes. Second, we remove the nodes without outgoing links in the remaining network and the incoming links incident to these nodes. The set of the nodes removed in the second round is denoted by $W_2$. We repeat the same procedure until all the nodes are removed. The set of the nodes removed in the $i$-th round is denoted by $W_i$. Now, a given DAG is assigned with a layer structure $\{W_1, W_2, ..., W_L\}$, where $L$ is the number of layers.

The rank of a node belonging to layer $W_i$ is given by $L-i+1$. The most and least dominant nodes possess ranks 1 and $L$, respectively. By construction, any directed link emanates from a worker with the higher rank to a worker with the lower rank. Multiple nodes may possess the same rank value. There is no link between nodes with the same rank (equivalently, nodes in the same layer).

**2.8 Generation of random DAGs**

We generate random DAGs that possess the same in-degree and out-degree of each node and the same distribution of the connected component size as those of the observed networks. The random DAGs are also used as null models for probing structure of observed networks. To generate networks, we use a previously proposed rewiring method (method c in [38]).

The rewiring process begins by applying the bottom-up leaf-removal algorithm to an observed network. Then, we randomly order the $N$ nodes in the network under the condition that a node with a higher rank (i.e., smaller rank value) appears earlier. This is tantamount to randomly ordering the nodes within each layer and align them from the last (i.e., $L$-th) to the first layer. Then, we select a directed link $k \to j$ (i.e., directed link from the $k$-th node to the $j$-th node) with the equal probability and then randomly select two nodes $i$ and $\ell$ such that (i) links $i \to k$ and $\ell \to j$ exist or (ii) links $k \to i$ and $j \to \ell$ exist. If condition (i) holds true, $\ell \to k$ does not exist, $i \to j$ does not exist, $\ell < k$ and $i < j$, then we replace $i \to k$ and $\ell \to j$ by $\ell \to k$ and $i \to j$, respectively. If condition (ii) holds true, $k \to \ell$ does not exist, $j \to i$ does not exist, $k < \ell$ and $j < i$, we replace $k \to i$ and $j \to \ell$ by $k \to \ell$ and $j \to i$, respectively. If any of these conditions is not satisfied, we repeat the same procedure with a different directed link $k \to j$ until a successful rewiring event occurs. Once we have rewired two links, we carry out the entire procedure starting from the leave-removal algorithm and iterate it until a desired number of links is rewired.

We verified that the generated DAG was random enough by measuring the so-called dissimilarity [38] (Electronic Supplementary Material).

**2.9 Out-strength**

The out-strength of a worker is defined as the sum of the weights of the links outgoing from the worker. It corresponds to the number of aggression behaviour that the worker has exerted on other workers and is identical to the *AttFr* dominance index [35].

**2.10 Reversibility**

The reversibility from non-maximal nodes (i.e., workers) to maximal nodes, denoted by $H$, quantifies the variety of paths in the DAG [39]. In a DAG, the maximal node is defined as the most dominant node, i.e., a node without incoming link. The $H$ value is the average amount of information necessary for reversely traveling from non-maximal to maximal nodes (see Electronic Supplementary Material for definition). If $H$ is equal to zero, the network is a (heterogeneous) directed tree, including the case of the directed chain, such that any subordinate worker is attacked by just one worker. A large $H$ value indicates that subordinate workers would often receive multiple incoming links and there tend to be various reversed pathways to reach one of the most dominant workers from a subordinate worker. The reversibility is a quantity exclusively defined for DAGs.

**2.11 Hierarchy index**

The hierarchy index denoted by $v$ ranges between $-1$ and $1$ and quantifies the extent to which the network has pyramidal structure and reversibility of paths [40] (see Electronic Supplementary Material for definition). A network with $v = 1$ is perfect (possibly heterogeneous) tree such that all nodes except a single root node have in-degree unity and all nodes except leaves, which possess out-degree zero, have out-degree at least two. In addition, the distance from the unique root node to any leaf node is the same. The network with $v = -1$ is an inverted (possibly heterogeneous) tree such that all nodes except a single leaf has out-degree unity and all nodes except roots, which possess in-degree zero, have in-degree at least two. In addition, the distance from any root to the unique leaf is the same. Networks having $v$ values close to zero are considered to lack hierarchical structure in both downward and upward directions. The hierarchical index is also a quantity exclusively defined for DAGs.

**2.12 Global reaching centrality**

The global reaching centrality *GRC* is defined as follows [41]. The local reaching centrality of node *i*, denoted by $C_R(i)$, is defined as the proportion of the other nodes reachable from node *i* by following outgoing links. On the basis of $C_R(i)$, we define

$$GRC = \frac{\sum_{i=1}^{N}\left[C_R^{\max} - C_R(i)\right]}{N-1}, \qquad (9)$$

where $C_R^{\max}$ is the maximum of $C_R(i)$ $(1 \leq i \leq N)$. *GRC* ranges between 0 and 1. It is an indicator similar to the CV in the meaning that it quantifies the heterogeneity of the out-degree. However, *GRC* also quantifies the level of hierarchy in a network. A large *GRC* value indicates that directed paths starting from a small fraction of nodes reach a majority of nodes such that the network has strong hierarchical structure. In particular, if the *GRC* value is equal to unity, the network is the directed outward star, in which a single hub sends a directed link to every other node. The *GRC* is a quantity defined for general directed networks including DAGs.

**2.13 Network motif**

Network motifs are overrepresented small subgraphs in a given network [42]. Out of the 13 possible directed and weakly connected three-node patterns, only four patterns (motifs 1, 2, 4, and 5 as defined in [42]) are possible in a DAG. It should be noted that here we are not concerned with the frequency of intransitive triads such as bidirectional links [24] and cycles [22, 25, 26] because our observed networks are (approximately) DAGs (see Results), which are devoid of intransitive triads. We calculate the number of each of the four three-node patterns in the observed networks, thinned linear tournaments, and randomised DAGs. We perform the motif analysis using the igraph package implemented in R.

**2.14 *Z* score**

To assess the significance of the quantities measured for the observed networks, such as *GRC*, we compare them with those calculated for null model networks, which are either thinned linear tournaments or random DAGs. To this end, we calculate the *Z* score, i.e., the distance between, for example, the *GRC* value for the observed network and the mean of *GRC* for the null model divided by the standard deviation of *GRC* for the null model. The mean and standard deviation for the null model are calculated on the basis of $10^3$ realisations of the null model. A large absolute value of the *Z* score implies that the observed network deviates from the null model in terms of the examined variable, e.g., *GRC*. The *Z* score is conventionally used in the motif analysis [42].

## 3. Results

**3.1 Observed dominance networks are perfect or approximate DAGs**

We observed dominance networks from six colonies. A directed link was assumed between two ants if aggressive behaviour between them was observed at least once during the recording period. We also counted the number of aggression in each interacting pair. In the following, we focus on the largest weakly connected component (i.e., connected component when the direction of the links is ignored) of each colony, which we refer to as the dominance network. In fact, the second largest weakly connected component in each colony contained at most two workers, such that it was negligible. The statistics of the six dominance networks is summarised in Table 1. Further statistics of the networks (modified Landau's *h* index [20], triangle transitivity metric [22], and total number of observed interactions) is shown in Table S1. The dominance network contained 24−38% of workers in each colony. The full information about the structure of the six networks and network-related properties of workers used in the following analysis are available as Electronic Supplementary Material.

We mainly analysed the unweighted directed network (where the number of attacks on each link was reduced to unity). All observed dominance networks apparently had hierarchical structure as shown in Fig. 1. They were sparse networks, i.e., only approximately 10% of pairs of workers among the possible pairs interacted, yielding large sparseness values (Table 1; the definition of sparseness is given in the table's caption). The triangle transitivity, a measurement of linearity suitable for sparse networks, was equal to unity for five of the six observed networks; they have perfect transitivity. It was equal to 0.96 for the other network (i.e., colony C5).

In fact, the five dominance networks were DAGs. Colony C5 was almost a DAG in the sense that it was a DAG if we removed two bidirectional links (Table 1; red thick lines in Figure 1(e)). Even this colony did not have any directed cycle of length larger than two, which contrasted to the results of previous studies showing the presence of some cycles in various dominance networks [22, 25, 26].

**3.2 Purely random sampling of links from a linear tournament does not explain observed dominance networks**

The complementary cumulative distribution of the in-degree (number of other workers that attack a given worker) and that of the out-degree (number of other workers that a worker attacks) are shown in Figure 2 (non-cumulative distributions are shown in Figure S3). For all colonies, both in-degree and out-degree were inhomogeneous among workers.

To quantify the heterogeneity in the degree, we measured its CV. We found that the CV values for the in-degree distribution for the observed dominance networks were much smaller than unity (Table 1); the in-degree was rather homogeneously distributed. In fact, most workers were attacked by just one or two other workers. In contrast, the CV for the out-degree distribution was much larger than unity in all colonies (Table 1). This result implies that some workers have dominated many others, and many workers have dominated few others. The results did not noticeably change upon the removal of the bidirectional links from C5 to make it a DAG (CV = 0.60 and 2.31 for the in- and out-degree, respectively).

Every DAG is consistent with the linear tournament, in which all pairs of individuals interact and are perfectly ranked such that the higher-ranked individual dominates the lower-ranked individual in any pair [1, 34]. However, the observed networks were sparse (i.e., small average degree and large sparseness; see Table 1). Therefore, we compared each observed networks with thinned linear tournaments having the same number of nodes and the same expected number of directed links as the observed network.

The CV value for both in-degree and out-degree predicted from the thinned linear tournament (Eq. (8)) was equal to 0.98 for C1, 0.93 for C2, 0.81 for C3, 0.87 for C4, 0.85 for C5, and 0.88 for C6. In short, the thinned linear tournament yielded CV values slightly smaller than unity. These CV values were consistently larger than those for the in-degree and much smaller than those for the out-degree for the observed networks. Therefore, the thinned linear tournament does not explain the observed dominance networks in terms of the CV.

**3.3 Origin of the heterogeneity in the frequency of aggression behaviour**

To further understand the origin of the heterogeneity of the out-degree, we classified the workers according to their ranks (determined by the bottom-up leaf-removal algorithm) in the hierarchy and calculated the mean out-degree of the workers having the same rank. In general, multiple workers may possess the same rank. Because this algorithm as well as the reversibility and hierarchy index analysed in the next section is exclusively applicable to DAGs, we preprocessed colony C5 by removing the two bidirectional links to make it a DAG for the present and following analysis.

The out-degree averaged over the workers with the same rank is plotted against the rank by the squares in Figure 3. A small rank value indicates a high rank in the hierarchy. The corresponding results for the thinned linear tournament are shown by the circles in Figure 3. The relationship between the out-degree and the rank was dissimilar between the observed networks (squares) and the thinned linear tournament (circles) even if we normalised the rank by the depth of the hierarchy, i.e., the total number of ranks. This result lends another support to our claim that the thinned linear tournament fails to explain the observed data. Figure 3 also indicates that the workers with disproportionately large out-degree values have a high but not the highest rank in all colonies.

The out-strength (i.e., the total number of aggression behaviour that the worker has exerted on other workers) averaged over the workers with the same rank is plotted against the rank in Figure 4. The figure indicates that workers near the top of the hierarchy have disproportionately large out-strength values. In some colonies, the averaged out-strength is the largest at the highest rank. In other colonies, the largest out-strength is realised by a high but not the highest rank, as is the case for the out-degree. To summarise the results shown in Figures 3 and 4, we conclude that workers near the top of the hierarchy have paid a disproportionately large amount of cost in attacking subordinates.

**3.4 Observed networks resemble randomised DAGs given the in-degree and out-degree of each worker**

In this section, we focus on unweighted dominance networks and examine the proximity of the observed networks, which are (approximate) DAGs, to thinned linear tournaments and randomised DAGs, using the following four types of analysis.

**Reversibility**: First, we measured the reversibility, $H$, to evaluate the variety of paths in the DAG. In all observed networks, the $H$ values were positive (Table 2). Then, we calculated the $Z$ score, i.e., distance between the $H$ value for the observed network and the mean of $H$ for the null model (either the thinned linear tournament or randomised DAG) divided by the standard deviation of $H$ for the null model. For three out of the six colonies, the $Z$ score was significantly positive when the null model was the thinned linear tournament. However, when the null model was the randomised DAG, the $Z$ score in four of the five colonies was close to zero such that the observed networks were not significantly different from the randomised DAGs (Table 2). It should be noted that, for colony C1, the $H$ value for any randomised DAG coincided with that for the observed network, making it impossible to calculate the $Z$ score.

We assumed that directed links emanated from attacking workers to attacked workers. If we adopt the opposite definition for the direction of links (i.e., from attacked to attacking), the $H$ value is generally altered. Therefore, we carried out the same statistical test for networks in which all links were reversed. The results were qualitatively the same as those for the original networks. In other words, the $Z$ score was significantly positive and insignificant when the null model was the thinned linear tournament and randomised DAG, respectively (Table S2).

**Hierarchy index:** Second, we measured the hierarchical index, $v$, to find that $v$ averaged over the six colonies was somewhat positive (mean ± SE: 0.30 ± 0.07). Therefore, the dominance networks had some hierarchical structure. For four colonies, $v$ was significantly larger for the observed networks than the thinned linear tournaments. However, for all colonies, $v$ was not significantly different from the value for the randomised DAG (Table 2).

Reversing the direction of all links in a given network only flips the sign of $v$. In addition, the link reversal of a thinned linear tournament is a thinned linear tournament. Therefore, the absolute values of the $Z$ score calculated for the link-reversed dominance networks were the same as those for the original dominance networks, up to statistical fluctuations (Table S2). In contrast, the randomised DAG is affected by the link reversal because a node generally has different in-degree and out-degree values. We confirmed that the results for the link-reversed networks with the randomised DAG as the null model were qualitatively the same as those for the original networks (Table S2). We conclude that, for a majority of observed networks, the hierarchical structure of the empirical networks is as expected from randomised DAGs.

**Global reaching centrality:** Third, we measured the global reaching centrality, $GRC$, to find that the $GRC$ value was large for all colonies (Mean ± SE; 0.86 ± 0.04), suggesting hierarchical structure. For all colonies, the $GRC$ value was significantly larger for the observed network than the thinned linear tournament. However, for all but one colony, the $GRC$ value was statistically indifferent from that for the randomised DAG (Table 2). The results were almost the same when we retained the bidirectional links in C5 ($GRC$ = 0.83). The results were also qualitatively the same when the same statistical test was applied to the dominance networks in which the direction of all links was reversed (Table S2).

**Network motif:** Fourth, we carried out motif analysis. The $Z$ scores for the four three-node patterns are shown in Fig. 5. The $Z$ score was significant for most three-node patterns when the null model was the thinned linear tournament. In contrast, the $Z$ score was insignificant in most cases when the null model was the randomised DAG.

Reversal of all links in a given network simply swaps motifs 1 and 4, and conserves motifs 2 and 5. Therefore, the link reversal conserved the $Z$ score when the null model was the thinned linear tournament except for statistical fluctuations and the swapping of the results for motif 1 and those for motif 4 (Figure S4). As shown in the same figure, the results for the link-reversed networks with the randomised DAG as the null model were qualitatively the same as those for the original networks.

To summarise the analysis of the four quantities in this section, we conclude that the randomised DAG, but not the thinned linear tournament, roughly approximates the observed dominance networks.

## 4. Discussion

We examined dominance networks formed by worker ants. By analysing the dominance networks as directed networks, we reached four main conclusions. First, the observed dominance networks are DAGs or approximately DAGs, which are consistent with perfect linear hierarchy despite their large sizes (Figure 1). Second, the out-degree obeys a much more heterogeneous distribution than the in-degree does (Figure 2 and Table 1). Third, the workers with high ranks showed a larger amount of aggressive behaviour than those with low ranks (Figures 3 and 4). Fourth, the dominance networks are indistinguishable from random DAGs under the condition that the in-degrees and out-degrees of all nodes are given (Figures 5 and S4, and Tables 2 and S2).

Empirical studies for various species often failed to detect perfect linear hierarchy in particular when networks were large [20–22, 25, 26]. In large groups, it would be impossible for individuals to recognise and interact with all other peers [19, 20, 22, 43]. The cost of exerting aggressive behaviour [44–46] may also contribute to the sparseness (i.e., low density of links) because the number of potential links per individual linearly increases with group size. For sparse networks, inference of the direction of missing links by previously established methods [19, 20] may be unreliable [22, 43, 47, 48]. To overcome this sparseness problem, the frequency of transitive triads (i.e., A dominates B, B dominates C, and A dominates C) has been used as an indicator of linear hierarchy [22, 25, 26, 49, 50]. In contrast, we analysed the dominance networks as DAGs. The triad census and our analysis methods are common in that they are suitable to large and sparse dominance networks. However, the triad census neglects emergent structure derived from a collection of more than three nodes. In contrast, we analysed global structure of networks to reveal that the observed networks were (approximate) DAGs. More importantly, our approach revealed the depth of hierarchy, hierarchical ranks of individual workers, and workers' behaviour depending on the rank. These parameters may be also informative in various biological and nonbiological contexts. In addition, the DAG analysis may be useful when dominance networks are very sparse and the triad census fails to detect significant orderliness of the network due to the paucity of triads. In fact, the $P$ value for the triangle transitivity metric [22] was not small enough for C1, C2, C5, and C6 (Table S1). This issue deserves further investigations.

The hierarchical ranking through the leaf-removal algorithm, redundancy, and hierarchy index are exclusively defined for DAGs. Therefore, we cannot immediately apply these methods to other dominance network data that we cannot transform to DAGs by removing just a few links, as we did for C5. Adapting the present methods to such data warrants future work.

It should be noted that the bottom-up leaf-removal algorithm for DAGs used in the present study produced multiple equal ranks, in particular for workers low in the hierarchy. In contrast, other established ranking methods produce a higher uniqueness of ranks, i.e., more different rank values for a given network [35, 51]. The former may be adequate when pairwise interaction occurs sparsely as in the present study such that the relative strength of many pairs of workers is unclear.

It has been discussed that linearity disappears in large dominance networks for the following proximate reason. In small groups, individuals can easily recognise each other and form a perfectly linear tournament even with a limited cognitive ability [21, 22]. In contrast, in large groups, as described above, it would not be possible for individuals to recognise and interact with all other peers [19, 20, 22, 43]. Our colonies contained up to 214 workers. However, the fraction of workers in the colony belonging to the dominance network, including purely subordinate workers, was not large, i.e., 24–38% (Table 1). In addition, the number of workers entering the hierarchy (i.e., showing aggressive behaviour) was small (5, 10, 11, 8, 18, and 16 workers in colonies C1, C2, C3, C4, C5, and C6, respectively) relative to the colony size. This number would even decrease if we only regard the workers showing frequent aggressive behaviour as entering the hierarchy [52]. The fact that only a small number of workers entered hierarchy and possibly had to recognise many others might contribute to the resultant nearly perfect linear hierarchy.

The present result that only a small fraction of workers has entered the hierarchy supports the prediction of inclusive fitness models [23, 53] in which workers are assumed to obtain direct fitness benefit by entering the hierarchy while suffer from indirect fitness costs. The prediction that a small fraction of workers enters the hierarchy is also consistent with empirical evidence for other ant species [14, 17, 52, 54] although the prediction quantitatively differs depending on intracolony relatedness and the offspring sex that workers directly produce [23, 53]. Furthermore, the same model [23, 53] predicts that the hierarchy is long for a large colony. This is intuitive because a worker in a large colony represents a small fraction of work force such that a worker joining the hierarchy without working does not much harm the colony. In the present study, the length of the hierarchy was operationally defined as the number of workers showing aggressive behaviour or the number of distinct ranks determined by the bottom-up leaf-removal algorithm. The latter quantity was equal to 4, 7, 8, 7, 11, and 8 for C1, C2, C3, C4, C5, and C6, respectively (squares in Figure 3). According to either definition of the hierarchy length, the present results are roughly consistent with the theoretical prediction. It should be noted that the observed networks possessed redundancy in terms of the density of links. The minimum number of directed links necessary to maintain a connected DAG is equal to $N-1$. In fact, each observed network had more than twice this number of directed links. Therefore, the observed short length of the hierarchy is not simply due to infrequent interactions.

We found new patterns in dominance networks of *Diacamma* sp. that were not anticipated. In particular, the out-degree was more heterogeneous than the in-degree. This result implies that a relatively few workers near the top of the hierarchy have attacked many workers, whereas many workers have not attacked any others, rendering themselves the most subordinate. The extent of heterogeneity was beyond that expected by the thinned linear tournament. High rankers may accept the cost of attacking because they have high chances to reproduce in this species [27, 31]. However, the high ranked, but not the highest ranked, workers had the largest out-degree on average (Figure 3). In three of the six colonies, this property held true even when we counted the total frequency of aggression per worker of a given rank (Figure 4). The most dominant workers tended to attack a relatively small number of workers, and the frequency of attacks on each of such subordinate workers was large. In contrast, the high-ranked but not the highest-ranked workers attacked many workers, and the number of attacks on each attacked worker was relatively small. The reason for this difference is unclear. Mathematical models may help explain the rationale behind this observation.

The observed dominance networks did not statistically differ from random DAGs given the in-degree and out-degree of each individual. In this sense, the observed networks may not be so complex as they apparently look. This result coincides with that for acyclic networks investigated in other domains such as citation networks [55]. The relative simplicity revealed in the present study paves the way to, above all, construction of new generative models for dominance networks and development of statistical procedures to interrogate structure of observed and artificial dominance networks. These tasks as well as clarifying the extent to which our results generalise to other species warrant future work.

The most important limitation of the present study is that we have constructed the dominance networks on the basis of the observation for four days. In fact, dominance networks may be dynamic, as has been reported for interaction networks of ants [56]. At the same time, the observed networks seem to grow in terms of the density of links as the observation time increases at least up to four days. Therefore, we should not overemphasise the sparseness of the observed networks. Clarification of this issue requires analysis methods for time-dependent networks [57] in addition to longer longitudinal data.

**Acknowledgments**

The authors thank Shigeto Dobata and Nobuyuki Kutsukake for valuable comments on the manuscript. We thank Keiichiro Higa, Teizo Shimabukuro, and Motobu Museum for collecting ants. Finally, we thank anonymous referee 2 for valuable and detailed feedback. HS and MSA are supported by Grant-in-Aid for JSPS Fellows (No. 11J03715 and 11J10448). NM acknowledges the support provided through Grants-in-Aid for Scientific Research (No. 23681039) from MEXT, Japan, CREST, JST and ERATO, Kawabayashi Large Graph Project, JST.

**Figure captions**

**Figure 1. Observed dominance networks.**
Each panel corresponds to a colony. The largest connected component is drawn for each colony. A circle represents a worker. The workers are aligned according to their hierarchical ranks as determined by the bottom-up leaf-removal algorithm. An arrow represents aggressive behaviour exerted by an attacking worker toward an attacked worker. The two bidirectional links in C5 are shown by the red thick bidirectional arrows.

**Figure 2. Complementary cumulative distributions of the in-degree and out-degree in dominance networks.**
The fraction of nodes with the in-degree and out-degree larger than or equal to the value shown on the horizontal axis is plotted for the six dominance networks. The solid and dashed lines correspond to the in-degree and out-degree, respectively.

**Figure 3. Dependence of the average out-degree on the worker's rank.**
The out-degree averaged over the workers possessing the same rank is plotted against the rank for each colony. The squares and circles represent the results for the observed dominance networks and the corresponding thinned linear tournament averaged over $10^3$ realisations, respectively. The error bars accompanying the circles represent the standard deviation.

**Figure 4. Dependence of the average out-strength on the worker's rank.**
The out-strength, i.e., the sum of the link weights over the outgoing links of a worker, averaged over the workers possessing the same rank is plotted against the rank for each colony.

**Figure 5. Results of the motif analysis.**
We calculated the $Z$ score for the frequency of each three-node network against each null model (i.e., thinned linear tournament or randomised DAG). Asterisks indicate significance levels (*: $p < 0.05$, i.e., $|Z| > 1.96$; **: $p < 0.01$, i.e., $|Z| > 2.58$).

**Table 1. Statistics of observed dominance networks.**
The size represents the number of workers in the colony. The number of nodes represents the number of workers contained in the largest weakly connected component. The other quantities shown in the table are calculated for the largest component. Apart from the largest component, there is a weakly connected component composed of two workers and a directed link between them in C1, C2, C3, and C6. Colonies C4 and C5 contain a unique weakly connected component. The sparseness is defined as the number of noninteracting pairs of workers divided by all possible pairs, i.e., $N(N-1)/2$ [22]. The sparseness ranges from 0 to 1, with 0 corresponding to the all-to-all network and 1 to the null network.

**Table 2. Statistical results for the reversibility, hierarchy, and global reaching centrality.**
For each index, we calculated the $Z$ score against each null model (i.e., thinned linear tournament or randomised DAG). The thinned linear tournament and randomised DAG are abbreviated as tournament and DAG in the table. Asterisks indicate significance levels (*: $p < 0.05$, i.e., $|Z| > 1.96$; **: $p < 0.01$, i.e., $|Z| > 2.58$).

Figure 1

C1 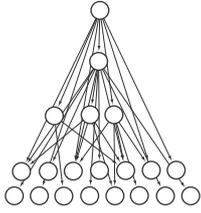

C2 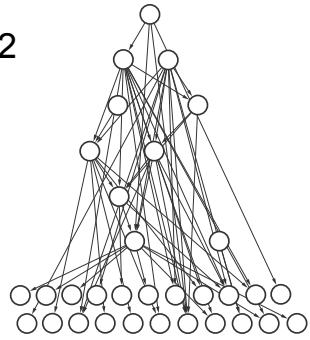

C3 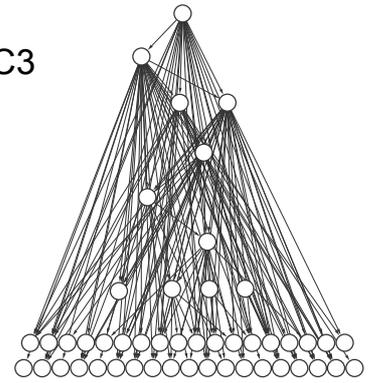

C4 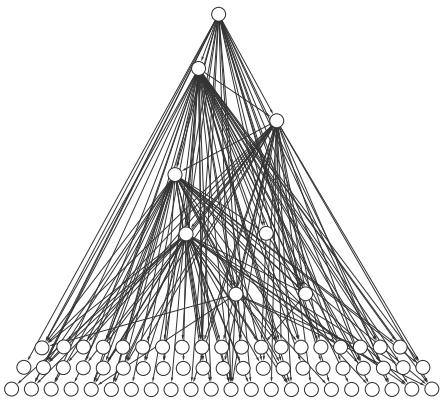

C5 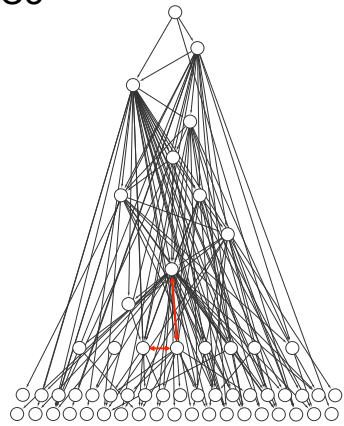

C6 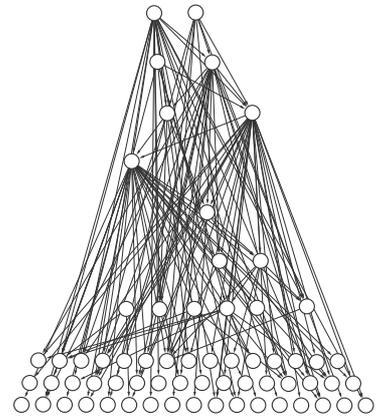

Figure 2

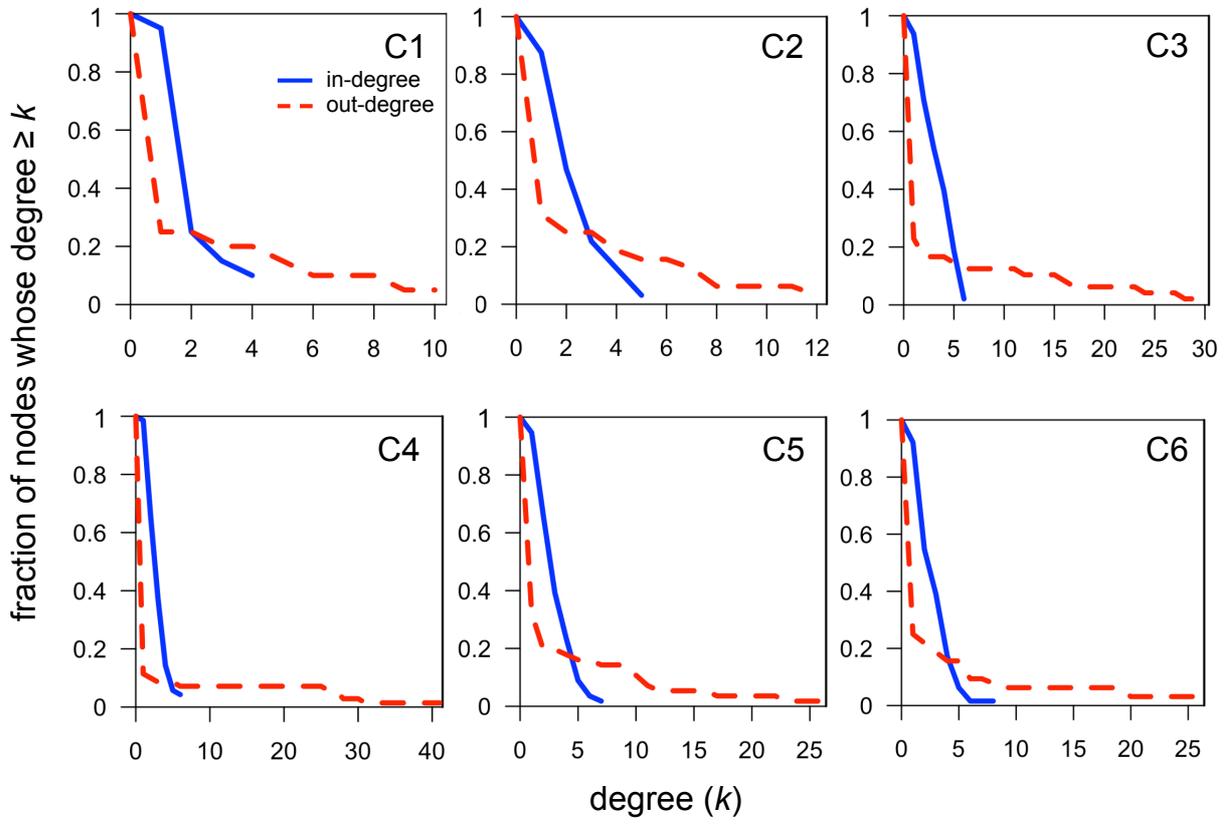

Figure 3

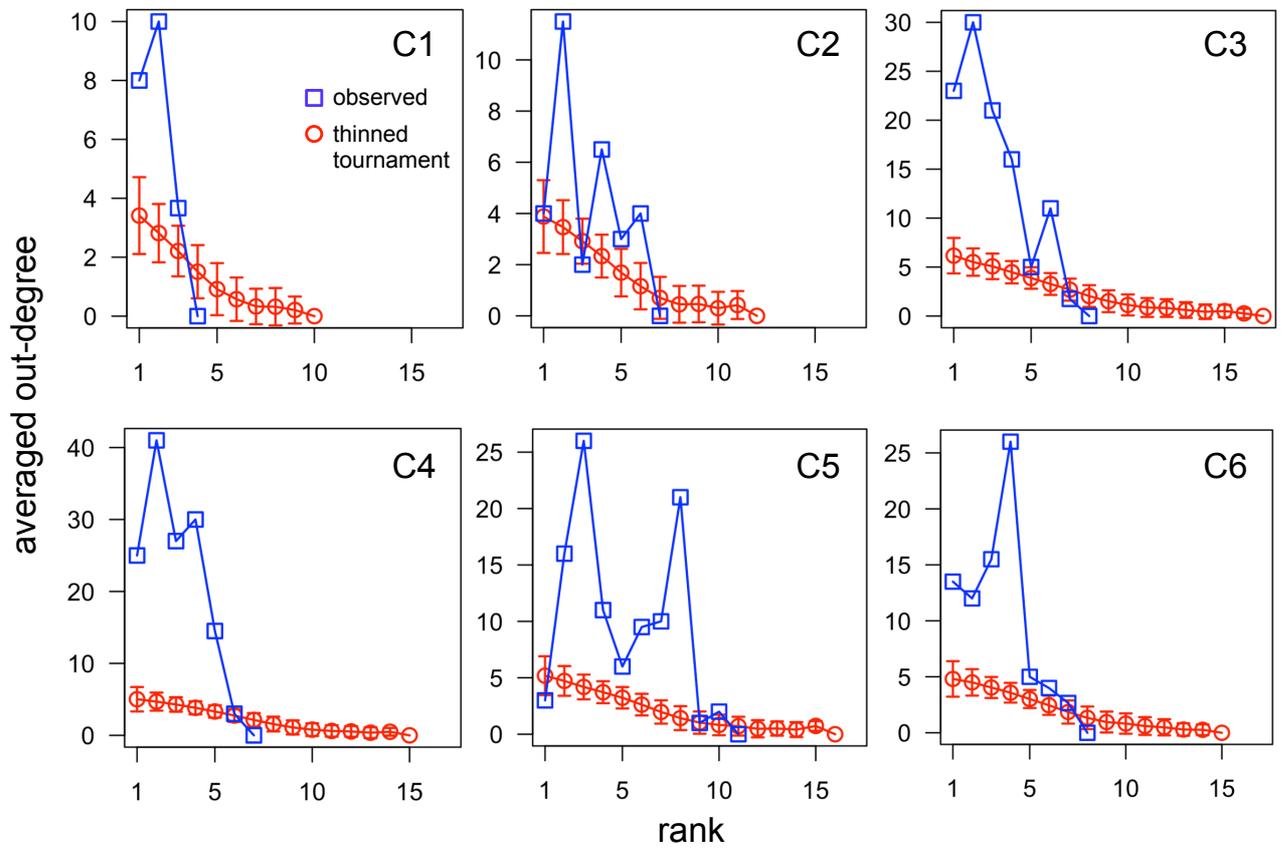

Figure 4

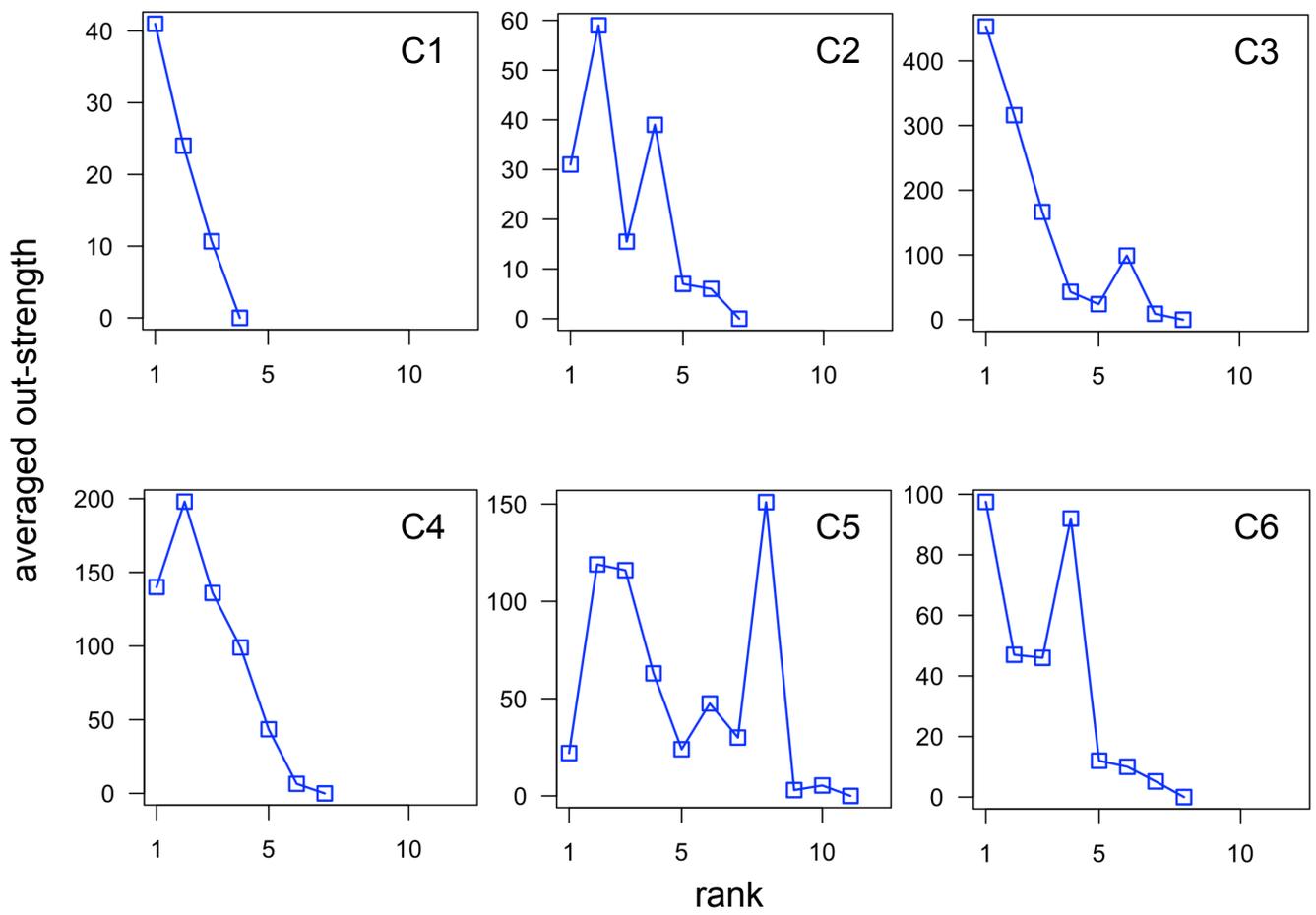

Figure 5

| Colony | Motif 1 | | Motif 2 | | Motif 4 | | Motif 5 | |
|---|---|---|---|---|---|---|---|---|
| | Thinned tournament | Randomised DAG | Thinned tournament | Randomised DAG | Thinned tournament | Randomised DAG | Thinned tournament | Randomised DAG |
| C1 | 6.64** | 0.08 | -0.99 | 0.08 | -1.46 | 0.08 | 0.99 | -0.08 |
| C2 | 7.42** | 0.93 | 1.23 | 0.93 | -2.20* | 0.93 | 4.85** | -0.93 |
| C3 | 25.91** | -2.14* | -3.31** | -2.14* | -4.81** | -2.14* | 13.65** | 2.14* |
| C4 | 49.23** | -0.45 | -1.77 | -0.45 | -5.25** | -0.45 | 22.80** | 0.45 |
| C5 | 18.73** | 0.99 | 2.95** | 0.99 | -4.00** | 0.99 | 13.93** | -0.99 |
| C6 | 24.67** | 1.82 | 1.91 | 1.82 | -3.32** | 1.82 | 14.22** | -1.82 |

Table 1

| Colony | Size | Number of nodes ($N$) | Number of links ($|E|$) | Average degree | CV of in-degree | CV of out-degree | Bidir. links | Sparseness |
|---|---|---|---|---|---|---|---|---|
| C1 | 58 | 20 | 29 | 2.90 | 0.71 | 1.99 | 0 | 0.85 |
| C2 | 132 | 32 | 55 | 3.44 | 0.73 | 1.90 | 0 | 0.89 |
| C3 | 149 | 48 | 134 | 5.58 | 0.58 | 2.55 | 0 | 0.88 |
| C4 | 183 | 70 | 158 | 4.51 | 0.58 | 3.49 | 0 | 0.93 |
| C5 | 200 | 56 | 133 | 4.75 | 0.64 | 2.29 | 2 | 0.91 |
| C6 | 214 | 64 | 137 | 4.28 | 0.71 | 2.63 | 0 | 0.93 |

Table 2

| Colony | Reversibility (*H*) | | | Hierarchy (*v*) | | | *GRC* | | |
|---|---|---|---|---|---|---|---|---|---|
| | value | Z score (tournament) | Z score (DAG) | value | Z score (tournament) | Z score (DAG) | value | Z score (tournament) | Z score (DAG) |
| C1 | 0.28 | −2.36* | − | 0.59 | 3.68** | −0.33 | 0.94 | 4.45** | 1.01 |
| C2 | 1.41 | 1.86 | 1.76 | 0.14 | 1.05 | −1.70 | 0.71 | 2.72** | −2.11* |
| C3 | 1.73 | 0.24 | 2.33* | 0.31 | 3.32** | 0.05 | 0.88 | 4.93** | −1.40 |
| C4 | 1.33 | −0.36 | −1.33 | 0.32 | 3.90** | −1.08 | 0.96 | 6.60** | 1.66 |
| C5 | 2.37 | 4.98** | 0.20 | 0.28 | 3.15** | 0.74 | 0.86 | 4.82** | −0.89 |
| C6 | 2.02 | 4.09** | 1.69 | 0.14 | 1.72 | 0.66 | 0.82 | 4.54** | −0.64 |

# Global network structure of dominance hierarchy of ant workers

Hiroyuki Shimoji, Masato S. Abe, Kazuki Tsuji, Naoki Masuda

## S.1 Randomness of rewired DAGs

The randomness of a DAG generated by the randomisation algorithm (article main text, section 2.6) depends on the number of successful rewiring events. To guarantee that we rewired the links sufficiently many times such that the generated DAG was random enough, we measured the dissimilarity [1] defined as

$$D = \frac{1}{2|E|} \sum_{i,j} \left[ 1 - \delta(A_{ij}, A'_{ij}) \right], \tag{S1}$$

where $A = (A_{ij})$ and $A' = (A'_{ij})$ were the adjacency matrices of the empirical network and the randomised network, respectively, and $\delta$ is the Kronecker delta. To obtain one random DAG from the degree sequence of an empirical network, we iterated the rewiring process until we successfully rewired the links 5000 times. The mean time courses of $D$ during the rewiring process are shown in Fig. S2. The figure shows that $D$ saturates sufficiently fast.

## S.2 Definition of reversibility

The reversibility $H$ is defined as follows [2]. In a DAG, the maximal node is defined as a node without incoming link, corresponding to the most dominant worker in the colony. There may be multiple maximal nodes in a DAG. We denote the set of all paths (consistent with the direction of links) from any maximal node to node $i$ by $\phi(i)$. Let $v(\pi_k)$ be the set of nodes participating in path $\pi_k \in \phi(i)$ except the maximal node.

$P(\pi_k | i)$ represents the probability of the reversed path of path $\pi_k$ starting from node $i$ under the unbiased random walk, i.e.,

$$P(\pi_k | i) = \prod_{j \in v(\pi_k)} \frac{1}{d_j^{\text{in}}}, \tag{S2}$$

where $d_j^{\text{in}}$ is the in-degree of node $j$. The uncertainty associated with the reversed paths starting from node $i$ is defined by

$$H(i) = -\sum_{\pi_k \in \phi(i)} P(\pi_k | i) \log P(\pi_k | i). \tag{S3}$$

The average of *H*(*i*) over all non-maximal starting nodes defines the reversibility as follows:

$$H = -\sum_{i \in V \setminus M} \frac{1}{N - |M|} \sum_{\pi_k \in \phi(i)} P(\pi_k | i) \log P(\pi_k | i). \tag{S4}$$

Here, *V*={1, 2,..., *N*} is the set of nodes, *M* is the set of the maximal nodes, and |·| denotes the number of nodes in the set. In practice, we can efficiently calculate *H* from the adjacency matrix [2].

**S.3 Definition of hierarchy**

To define the hierarchy index *v* [3], we first define the minimal node as a node without any outgoing link, corresponding to the most subordinate workers in the colony. The set of the minimal nodes is denoted by *μ*. Similar to equation (S4), the uncertainty value averaged over the minimal nodes, which are used as the starting nodes of paths, is equal to

$$H(\mu) = -\sum_{i \in \mu} \frac{1}{|\mu|} \sum_{\pi_k \in \phi(i)} P(\pi_k | i) \log P(\pi_k | i). \tag{S5}$$

The degree of pyramidal structure is defined by

$$H^{\text{fwd}}(M) = -\sum_{i \in M} \frac{1}{|M|} \sum_{\pi_k \in \phi^{\text{fwd}}(i)} P^{\text{fwd}}(\pi_k | i) \log P^{\text{fwd}}(\pi_k | i) \tag{S6}$$

and

$$P^{\text{fwd}}(\pi_k | i) = \prod_{j \in v^{\text{fwd}}(\pi_k)} \frac{1}{d_j^{\text{out}}}, \tag{S7}$$

where $\phi^{\text{fwd}}(i)$ is the set of all forward paths from *i* to a minimal node, $v^{\text{fwd}}(\pi_k)$ is defined as the set of the nodes participating in path $\pi_k \in \phi^{\text{fwd}}(i)$ except the minimal node, and $d_j^{\text{out}}$ is the out-degree of node *j*. We define the balance between $H(\mu)$ and $H^{\text{fwd}}(M)$ as

$$f(G) = \frac{H^{\text{fwd}}(M) - H(\mu)}{\max\{H^{\text{fwd}}(M), H(\mu)\}}. \tag{S8}$$

For a later use, we explicitly indicated that *f* is a function of a given network *G* in equation (S8). The denominator is the normalisation factor to ensure $-1 \leq f(G) \leq 1$.

The hierarchical index *v* is defined as the *f* value averaged over the networks obtained in the course of two leaf-removal algorithms. It is given by

$$v = \frac{1}{2L-3}\left\{f(G) + \sum_{i<L-1}\left[f(G_i) + f(\tilde{G}_i)\right]\right\}, \tag{S9}$$

where $G_i$ and $\tilde{G}_i$ are the networks obtained after the application of the first $i$ rounds of the bottom-up leaf-removal algorithm and the top-down leaf-removal algorithm (i.e., successively removing the nodes without incoming links) to the original network $G$, respectively. It should be noted that the number of layers, $L$, is the same for the bottom-up and top-down leaf-removal algorithms. The hierarchy index $v$ ranges from $-1$ to $1$. When $v$ is close to $1$, the network has pyramidal structure and needs little information for walking backward from the minimal nodes to maximal nodes. In other words, most nodes of the network have one incoming link and multiple outgoing links. Conversely, when $v$ is close to $-1$, most nodes have multiple incoming links and one outgoing link.

**S.4 Modified Landau's h index and triangle transitivity metric**

To calculate two orderliness measures for dominance networks, we treat the observed networks as unweighted networks in this section. We regard the two bidirectional links in C5 as unidirectional links by discarding one of the two directions whose link weight is smaller than the other [4, 5].

The modified Landau's $h$ index, denoted by $h'$, is calculated as follows [5, 6]. First, we fill all the missing dyads randomly. In other words, we assume a unidirectional link in either direction with probability $1/2$ for all node pairs between which a link is originally absent. The obtained network, which we call the imputed network according to [5], is a directed complete graph (also called tournament). Second, we calculate

$$h \equiv \frac{12}{N^3 - N}\sum_{i=1}^{N}\left(d_i^{out} - \frac{N-1}{2}\right)^2, \tag{S10}$$

where $d_i^{out}$ is the out-degree of node $i$ in the imputed network. $h$ ranges from $0$ to $1$, and the perfectly linear tournament yields $h = 1$. Finally, $h'$ is the average of $h$ over $10^4$ independent imputed networks. To calculate the $P$ value for $h'$ as defined in [5, 6], we compare each of the $10^4$ values of $h$ used for calculating $h'$ and the $h$ value calculated for a randomised imputed network. We generate a randomised imputed network by independently reassigning one of the two directions with probability $1/2$ to each link in the imputed network, which is the complete graph. To make the comparison $10^4$ times, we generate $10^4$ independent randomised imputed networks. Then, we count the fraction of times out of the $10^4$ times in which $h$ for the randomised network is larger than or equal to that obtained from the original imputed network. This fraction defines the one-tailed $P$ value for $h$'.

The triangle transitivity metric $t_{tri}$ is defined by

$$t_{tri} = 4\left(\frac{N_{transitive}}{N_{transitive} + N_{cycle}} - 0.75\right), \tag{S11}$$

where $N_{transitive}$ and $N_{cycle}$ are the frequencies of transitive triads (i.e., A dominates B, A dominates C, and B dominates C) and cyclic triads (i.e., A dominates B, B dominates C, and C dominates A) in the given dominance network, respectively [5]. A network in which all triangles are transitive yields $t_{tri} = 1$. A random network yields $t_{tri} \approx 0$. To calculate the one-tailed $P$ value for $t_{tri}$, we generate $10^3$ random directed networks possessing the same numbers of nodes and links as those of the observed network, calculate $t_{tri}$ for each random network, count the number of the random networks yielding $t_{tri}$ values larger than or equal to that obtained from the observed network, and divide the count by $10^3$.

The values of $h$', $t_{tri}$, and their $P$ values for the six colonies are shown in Table S1. It should be noted that these quantities are measured for the largest weakly connected component of each colony, whose size is shown in Table 1. The total number of observed interactions, including those observed in the small components, is also shown in the table. Table S1 indicates that the $P$ values for $h$' and $t_{tri}$ are larger than or equal to 0.05 (i.e., no significant linearity) except for colonies C3 and C4 (and C5 in case of $t_{tri}$). It should be noted that the $P$ value for $t_{tri}$ is large although $t_{tri}$ for the observed network takes the maximum possible value (i.e., =1) in C1, C2, and C6. This is because randomised networks often yield $t_{tri} = 1$ owing to the sparseness of the network.

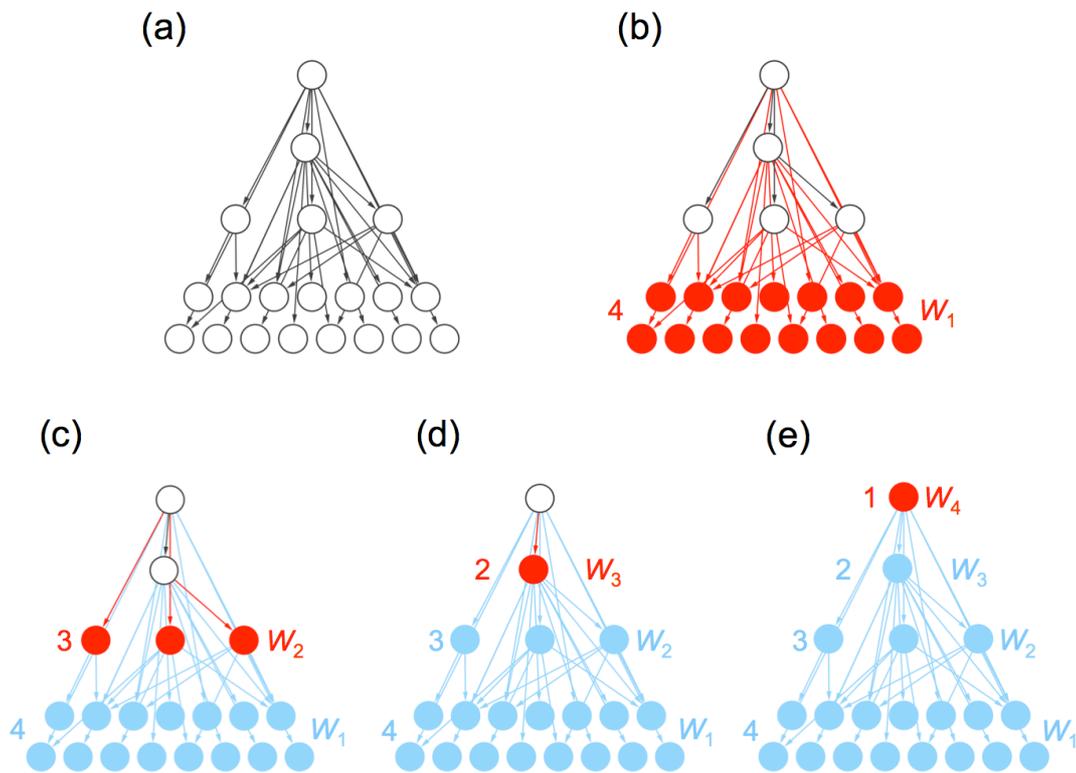

**Figure S1. Bottom-up leaf-removal algorithm.**

(a) Colony C1, replicating the left-top panel in Figure 1. In the first step of the algorithm, the nodes shown in red in (b) are removed because their out-degree is equal to zero. These nodes form layer $W_1$ and in fact receive rank value 4 because it turns out at the end of the algorithm that there are four layers. We also remove the links incident to the removed nodes. Second, the three nodes shown in red in (c) are removed because their out-degree is zero in the reduced network. These nodes form layer $W_2$ and receive rank value 3. Third, the node shown in red in (d), which constitutes layer $W_3$ and receives rank value 2, is removed. Finally, the node shown in red in (e), which constitutes layer $W_4$ and receives rank value 1, is removed.

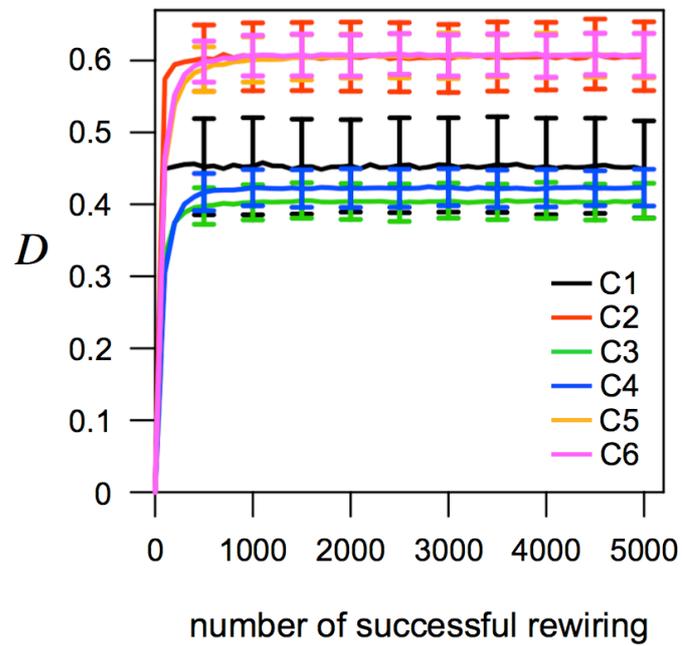

**Figure S2. Time courses of dissimilarity during the rewiring process.**
Each curve represents the time course of the dissimilarity (i.e., $D$) for one colony averaged over $10^3$ realisations of the randomisation runs, each starting from the observed network. The error bars represent standard deviations.

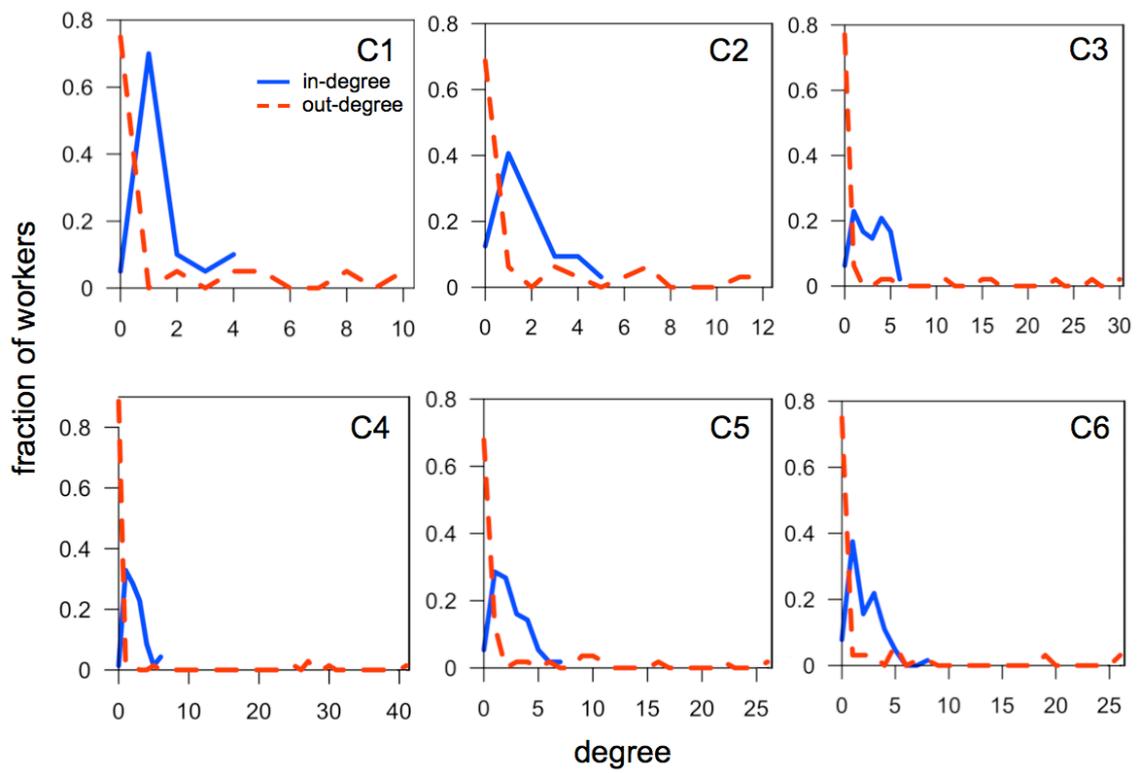

**Figure S3. Distributions of the in-degree and out-degree in dominance networks of colonies C1 to C6.**

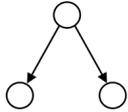

| Colony | Motif 1 | | Motif 2 | | Motif 4 | | Motif 5 | |
|---|---|---|---|---|---|---|---|---|
| | Thinned tournament | Randomised DAG | Thinned tournament | Randomised DAG | Thinned tournament | Randomised DAG | Thinned tournament | Randomised DAG |
| C1 | -1.40 | -0.21 | -0.96 | -0.21 | 6.84** | -0.21 | 1.02 | 0.21 |
| C2 | -2.11* | 0.67 | 1.12 | 0.67 | 7.17** | 0.67 | 4.73** | -0.67 |
| C3 | -4.96** | -2.35* | -3.23** | -2.35* | 25.77** | -2.35* | 13.79** | 2.35* |
| C4 | -5.39** | -0.51 | -1.78 | -0.51 | 51.36** | -0.51 | 22.90** | 0.51 |
| C5 | -4.04** | 0.51 | 3.05** | 0.51 | 18.59** | 0.51 | 13.41** | -0.51 |
| C6 | -3.29** | 0.59 | 1.90 | 0.59 | 24.35** | 0.59 | 14.37** | -0.59 |

**Figure S4. Results of the motif analysis for the link-reversed networks.** For the networks generated by reversing all links in the observed dominance networks, we calculated the $Z$ score for the frequency of each motif against each null model (i.e., thinned linear tournament or randomised DAG). Asterisks indicate significance levels (*: $p < 0.05$, i.e., $|Z| > 1.96$; **: $p < 0.01$, i.e., $|Z| > 2.58$).

**Table S1. Other statistics of observed dominance networks.**

The modified Landau's h index, h', triangle transitivity metric $t_{tri}$, their P values, and the total number of observed interactions for each colony are shown. We calculated the number of interactions for the entire colony and the other quantities for the largest weakly connected component of the colony.

| Colony | h' | P value for h' | $t_{tri}$ | P value for $t_{tri}$ | Number of interactions |
|--------|------|----------------|-----------|------------------------|------------------------|
| C1 | 0.21 | 0.18 | 1.00 | 0.39 | 98 |
| C2 | 0.12 | 0.23 | 1.00 | 0.23 | 278 |
| C3 | 0.13 | 0.0003 | 1.00 | 0.001 | 1306 |
| C4 | 0.08 | 0.0005 | 1.00 | 0.029 | 673 |
| C5 | 0.07 | 0.09 | 0.96 | 0.024 | 680 |
| C6 | 0.07 | 0.05 | 1.00 | 0.053 | 537 |

**Table S2. Statistical results for the reversibility, hierarchy, and global reaching centrality for the dominance networks with reversed links.**

See the caption of Table 2 for the legends.

| Colony | Reversibility ($H$) | | | Hierarchy ($v$) | | | GRC | | |
|---|---|---|---|---|---|---|---|---|---|
| | value | Z score (tournament) | Z score (DAG) | value | Z score (tournament) | Z score (DAG) | value | Z score (tournament) | Z score (DAG) |
| C1 | 1.76 | 3.84** | −0.78 | −0.59 | −3.41** | 0.19 | 0.16 | −2.23* | 1.10 |
| C2 | 2.21 | 5.12** | −0.05 | −0.14 | −1.04 | −0.36 | 0.14 | −2.88** | −2.43* |
| C3 | 2.13 | 2.09* | −0.66 | −0.31 | −3.32** | −0.98 | 0.08 | −5.83** | −1.29 |
| C4 | 2.57 | 6.87** | −0.40 | −0.32 | −3.91** | −0.41 | 0.05 | −5.09** | −1.10 |
| C5 | 1.92 | 2.75** | −2.42* | −0.28 | −3.17** | −3.13** | 0.12 | −4.04** | −1.39 |
| C6 | 2.06 | 4.21** | −1.26 | −0.14 | −1.59 | −2.23* | 0.11 | −4.08** | −1.28 |

**Supplementary References**

**Supplementary Files**

The following files include the nodes that do not belong to the largest weakly connected component.

**List of links:** C1.txt, ... , C6.txt available as the Electronic Supplementary Material contain the list of links in the corresponding colony. In each file, each row contains three columns. The first column represents the attacking worker's ID. The worker's ID starts with 1. The second column represents the attacked worker's ID. The third column represents the link weight, i.e., the frequency of attacks observed for the pair. Colonies C1, C2, C3, and C6 contain the second largest DAG composed of a single directed link in addition to the largest DAG (i.e., largest weakly connected component) analysed in the present study. Such a single link is from 18 to 3 in C1, from 5 to 11 in C2, from 48 to 42 in C3, and from 23 to 44 in C6. Colony C5 contains two bidirectional links, one between 5 and 42, and the other between 12 and 42.

**Node properties:** node_property.txt available as Electronic Supplementary Material contains the list of nodes with their properties. Each row contains six columns, i.e., worker's ID, out-degree, in-degree, out-strength, in-strength, and rank as determined by the bottom-up leaf-removal algorithm (from the first to sixth columns). The worker's IDs are the same as those used in C1.txt, ... , C6.txt. The data for each colony starts by a row reading "[C1]", "[C2]", and so on.